\documentclass[journal]{IEEEtran}
\usepackage{amsmath,amsfonts}
\usepackage{algorithmic}
\usepackage{algorithm}
\usepackage{array}
\usepackage[caption=false,font=normalsize,labelfont=sf,textfont=sf]{subfig}
\usepackage{textcomp}
\usepackage{stfloats}
\usepackage{url}
\usepackage{verbatim}
\usepackage{graphicx}
\usepackage{cite}
\usepackage{tikz}
\usepackage[T1]{fontenc}
\usepackage{authblk}

\begin{document}

\title{Evaluating tDCS Intervention Effectiveness via Functional Connectivity Network on Resting-State EEG Data  in Major Depressive Disorder}

\author[1]{Vishwani Singh}
\author[2]{Rohit Verma}
\author[3]{Shaurya Shriyam}
\author[4]{Tapan K. Gandhi*}

\affil[1]{Khoury College of Computer Sciences, Northeastern University, Boston, USA}
\affil[2]{Dept. of Psychiatry, All India Institute of Medical Sciences Delhi, India}
\affil[3]{Dept. of Mechanical Engineering, Indian Institute of Technology Delhi, India}
\affil[4]{Dept. of Electrical Engineering, Indian Institute of Technology Delhi, India}

\markboth{Journal of XXX,~Vol.~Y, No.~Z, April~2024}%
{Shell \MakeLowercase{\textit{et al.}}: A Sample Article Using IEEEtran.cls for IEEE Journals}

\maketitle

\begin{abstract} 
Transcranial direct current stimulation (tDCS) has emerged as a promising non-invasive therapeutic intervention for major depressive disorder (MDD), yet its effects on neural mechanisms remain incompletely understood. This study investigates the impact of tDCS in individuals with MDD using resting-state EEG data and network neuroscience to analyze functional connectivity. We examined power spectral density (PSD) changes and functional connectivity (FC) patterns across theta, alpha, and beta bands before and after tDCS intervention. A notable aspect of this research involves the modification of the binarizing threshold algorithm to assess functional connectivity networks, facilitating a meaningful comparison at the group level. Our analysis using optimal threshold binarization techniques revealed significant modifications in network topology, particularly evident in the beta band, indicative of reduced randomization or enhanced small-worldness after tDCS.
Furthermore, the hubness analysis identified specific brain regions, notably the dorsolateral prefrontal cortex (DLPFC) regions across all frequency bands, exhibiting increased functional connectivity, suggesting their involvement in the antidepressant effects of tDCS. Notably, tDCS intervention transformed the dispersed high connectivity into localized connectivity and increased left-sided asymmetry across all frequency bands. Overall, this study provides valuable insights into the effects of tDCS on neural mechanisms in MDD, offering a potential direction for further research and therapeutic development in the field of neuromodulation for mental health disorders.
\end{abstract}



\section{Introduction}

\IEEEPARstart{M}{}ajor depressive disorder (MDD) is a mental illness marked by an enduring state of melancholy, guilt, worthlessness, and hopelessness, which may increase the risk of suicidal behavior\cite{jia2010high}. MDD is the most common type of unipolar affective disorder, characterized by abnormal brain activity\cite{smith2009serotonin}. According to data released by the World Health Organisation (WHO), over $350$ million individuals worldwide suffer from MDD and the number of new cases has increased by almost $18\%$ over the past ten years\cite{peng2019multivariate}.


tDCS involves the application of low-intensity electrical currents to specific brain regions via scalp electrodes, resulting in the modulation of neuronal excitability and synaptic activity\cite{nitsche2008transcranial}. Unlike other neuromodulation techniques, such as selective serotonin reuptake inhibitors\cite{frank2009treatment}, tDCS is portable, well-tolerated, and relatively inexpensive, making it an attractive option for both research and clinical applications.

tDCS is considered for treating MDD because of evidence suggesting problems in neural circuitry (abnormalities in the connections between different brain regions) and neurotransmitter dysregulation (imbalance in the levels or activity of neurotransmitters) contributing to depression. Neuroimaging studies have consistently identified alterations in brain structure and function, particularly within corticolimbic circuits involved in emotional regulation and mood processing, in individuals with MDD\cite{drevets2008neuroimaging}. Dysfunction within these circuits, including the dorsolateral prefrontal cortex (DLPFC)-limbic system pathway, has been implicated in developing and maintaining depressive symptoms\cite{mayberg1997limbic, mayberg2003modulation}.

The antidepressant effects of tDCS in MDD have been investigated in numerous clinical trials and meta-analyses, with accumulating evidence supporting its efficacy and tolerability as a standalone treatment or adjunctive therapy\cite{brunoni2016clinical}. Additionally, studies have explored the potential of tDCS to induce sustained antidepressant effects through repeated or maintenance treatments, as well as its combination with other therapeutic modalities to enhance treatment outcomes\cite{brunoni2014treatment}.

As research into transcranial Direct Current Stimulation (tDCS) for Major Depressive Disorder (MDD) continues to expand, several critical challenges remain. These include optimizing stimulation parameters, elucidating underlying neurobiological mechanisms, and identifying predictors of treatment response. Addressing these challenges is essential for advancing our understanding of tDCS as a therapeutic intervention for MDD and enhancing its clinical efficacy.

This paper aims to contribute to our understanding of the alterations in neural circuitry induced by tDCS and its potential role in the management of depression.


\section{Related works}

Functional connectivity analysis and network neuroscience are valuable tools in neuroimaging for understanding brain function. tDCS has gained interest for its potential in treating MDD. Though its exact mechanisms are complex, research has highlighted key neurobiological processes involved in tDCS-induced antidepressant effects. Functional connectivity examines statistical associations between brain regions, revealing networks underlying cognition and emotion. Network neuroscience characterizes brain networks mathematically, assessing properties such as average path length and clustering. In depression-related research, these approaches have shown altered connectivity and network organization, where tDCS has been found to modulate these networks.

Nitsche et al. (2008) proposed that anodal stimulation, by promoting neuronal depolarization, enhances synaptic strength and induces long-term potentiation (LTP) in targeted brain regions, and cathodal stimulation, by decreasing neuronal depolarization, induces long-term depression (LTD) and reduces cortical excitability \cite{nitsche2008transcranial}. This modulation of neural activity within mood-relevant circuits is thought to restore dysfunctional neural circuits associated with MDD and alleviate depressive symptoms.

Animal studies have been instrumental in elucidating the neuroplasticity-related changes induced by tDCS. For example, Monai et al. (2016) demonstrated that tDCS leads to enduring alterations in synaptic plasticity and neurotransmitter levels, particularly within the glutamatergic and gamma-aminobutyric acid (GABA) systems \cite{monai2016brain}. These findings suggest that tDCS may exert its therapeutic effects by modulating neuronal excitability and synaptic strength in specific brain regions implicated in mood regulation. 

Research on functional connectivity in MDD has revealed several consistent findings. Mulders et al. (2015) identified increased connectivity within the anterior default mode network, between the salience network and the anterior default mode network, and decreased connectivity between the posterior default mode network and the central executive network \cite{mulders2015resting}. Fattahi et al. (2021) further explored these changes, finding significant differences in connectivity between various network pairs in MDD patients with suicidal thoughts \cite{fattahi2021functional}. Zhi et al. (2018) identified the disrupted topological organization of dynamic functional network connectivity in MDD, with patients spending more time in a weakly connected state associated with self-focused thinking \cite{zhi2018aberrant}.

The brain is an extremely complex multivariate dynamical network. Network neuroscience employs mathematical tools to model these intricate network systems using the concept of nodes (also called vertices) and edges\cite{biggs1999graph}. In the context of EEG connectivity, nodes represent electrodes, while edges represent the connectivity between electrode pairs. Thus, brain networks and brain functional connectivity are depicted in either matrix or graph form.

Past studies have highlighted the application of graph theory in understanding brain networks. Chung (2021) and Vecchio et al. (2017) both explored the scale-free and small-world properties of these networks, with the latter also discussing the potential for graph theory to aid in understanding brain disconnection and monitoring treatment impact \cite{chung2021graph, vecchio2017connectome}. Additional studies have underscored the significance of graph theory in unravelling the architecture, development, and evolution of brain networks\cite{fallani2012graph, sporns2018graph}.

Studies of resting-state EEG data in MDD using functional connectivity and graph theoretical approaches suggest that depressed patients exhibit altered brain functional connectivity patterns compared to healthy individuals. These studies highlight the potential of multilayer brain functional network frameworks for analyzing abnormal brain interaction patterns in depression\cite{sun2023study}. Furthermore, the effect of tDCS in modulating resting-state functional connectivity, particularly in regions distal to the stimulation site, has been emphasized\cite{bouchard2023changes}. Research utilizing graph theory to analyze brain networks in MDD has revealed several key findings. Hasanzadeh et al. (2020) and Hasanzadeh et al. (2017) both found that MDD patients exhibit a more randomized structure in their brain networks, with higher node degree and strength \cite{hasanzadeh2020graph, hasanzadeh2017investigation}. Sun et al. (2019) further confirmed these findings by identifying deficiencies in right hemisphere function and a randomized network structure in MDD \cite{sun2019graph}. Ye et al. (2015) added to this finding by demonstrating that MDD patients have higher local efficiency and modularity in their brain networks, as well as altered nodal centralities in specific brain regions, implying that emotional and cognitive functions are significantly affected by MDD \cite{ye2015changes}. 

These insights support the hypothesis that statistically significant differences exist in binarized network structures in MDD. These differences are quantified using network neuroscience measures such as small-worldness, hubness, and asymmetry. This approach provides a promising avenue for further exploration and understanding of MDD pathology. 


\section{Data Collection}

Individuals diagnosed with MDD were administered tDCS to manage symptoms. Resting-state EEG data were recorded before and after the intervention.

\subsection{Research design}
\begin{list}{}{}
\item \textbf{Study type:} Retrospective Data Analysis
\item \textbf{Study site:} Brain Stimulation Facility, Department of Psychiatry, All India Institute of Medical Sciences (AIIMS), New Delhi.
\end{list}

\subsection{Sampling method}
\indent All participants diagnosed with MDD according to the Diagnostic and Statistical Manual of Mental Disorders version $5$ (DSM $5$), who completed 20 sessions of tDCS and had pre- and post- EEG datasets available were included in this study. Eligibility for tDCS intervention required the absence of pregnancy, metal implants, or electronic devices. Written informed consent was obtained from all participants before the initiation of the treatment. Participants continued their ongoing antidepressant medications, if any, at the same dosage throughout the intervention period.

\subsection{tDCS intervention}
\indent The high-definition tDCS (HD-tDCS) intervention was administered using a battery-driven, constant-current stimulator (``Starstim R3'' system, Neuroelectrics, Barcelona, Spain). A current of $2 mA$ was delivered for $20$ minutes per session, with 30 seconds ramp-up and ramp-down. Participants received two sessions per day with a minimum interval of $3$ hrs between sessions, totalling $20$ sessions over $10$ days. The anode was placed at the F3 location, and the cathodal returns were positioned in a ring around the anode (FC1, AF3, F7, FC5).

\subsection{Procedure}

\textbf{Screening and recruitment:} Potential participants were screened to assess eligibility criteria. 

\textbf{Questionnaires:} Participants' demographic profiles were collected, and illness severity was assessed using the Hamilton Depression Rating Scale (HAMD). The HAMD is a 17-item clinical questionnaire focusing on physical and melancholic symptoms of depression. Depression severity was categorized as follows: $0-7$ (normal), $8-16$ (mild depression), $17-23$ (moderate depression) and $\geq24$ (severe depression)\cite{zimmerman2013severity}.

\textbf{EEG data collection:} EEG data were collected at baseline (pre-intervention) and post-intervention for both groups using standardized procedures with the Starstim R32 system. During the recording of $32$ channels, two ASCII files were generated with the \textit{\*.info} and \textit{\*.easy} file extensions. The \textit{*.info} file consists of metadata such as sampling rate and channel positions, while the \textit{\*.easy} file included voltage measurements (in $nV$), three acceleration columns (in $mm/s^2$), a trigger column and a timestamp column. The device settings included a sampling rate of $500$ samples/sec, with line filtering enabled to reduce power line interference ($50-60 Hz$), while FIR filter, EOG correction filter and reference filter were disabled.

\subsection{Data collection timeline}
Data were collected at three key points concerning the tDCS intervention:
\begin{list}{}{}
\item \textbf{Pre-intervention data (at baseline):} This dataset included demographic profiles, HAMD assessments and a 5-minute recording of resting-state, eye-closed EEG data.
\item \textbf{Post-intervention data (after 20 tDCS sessions):} This dataset comprised HAMD assessments and a 5-minute recording of resting-state, eye-closed EEG data.
\item \textbf{Follow-up data (after 4 weeks of last tDCS session):} This dataset included only HAMD assessments.
\end{list}

\subsection{Data analysis}
Intervention efficacy was assessed by examining the differences in HAMD scores at baseline, after $20$ tDCS sessions and at follow-up after weeks using a repeated measures analysis of variance (rmANOVA). EEG data were analyzed using network analysis techniques described later in the paper to compare functional connectivity between groups.

\subsection{Ethical considerations}
This study received ethical approval from the AIIMS Institute Ethics Committee (IEC-$169/04.03.2022$, RP-$14/2022$). Written informed consent was obtained from all participants, and their confidentiality and privacy were strictly maintained.

\begin{figure*}[!t]
  \centering
  \includegraphics[width=1\linewidth]{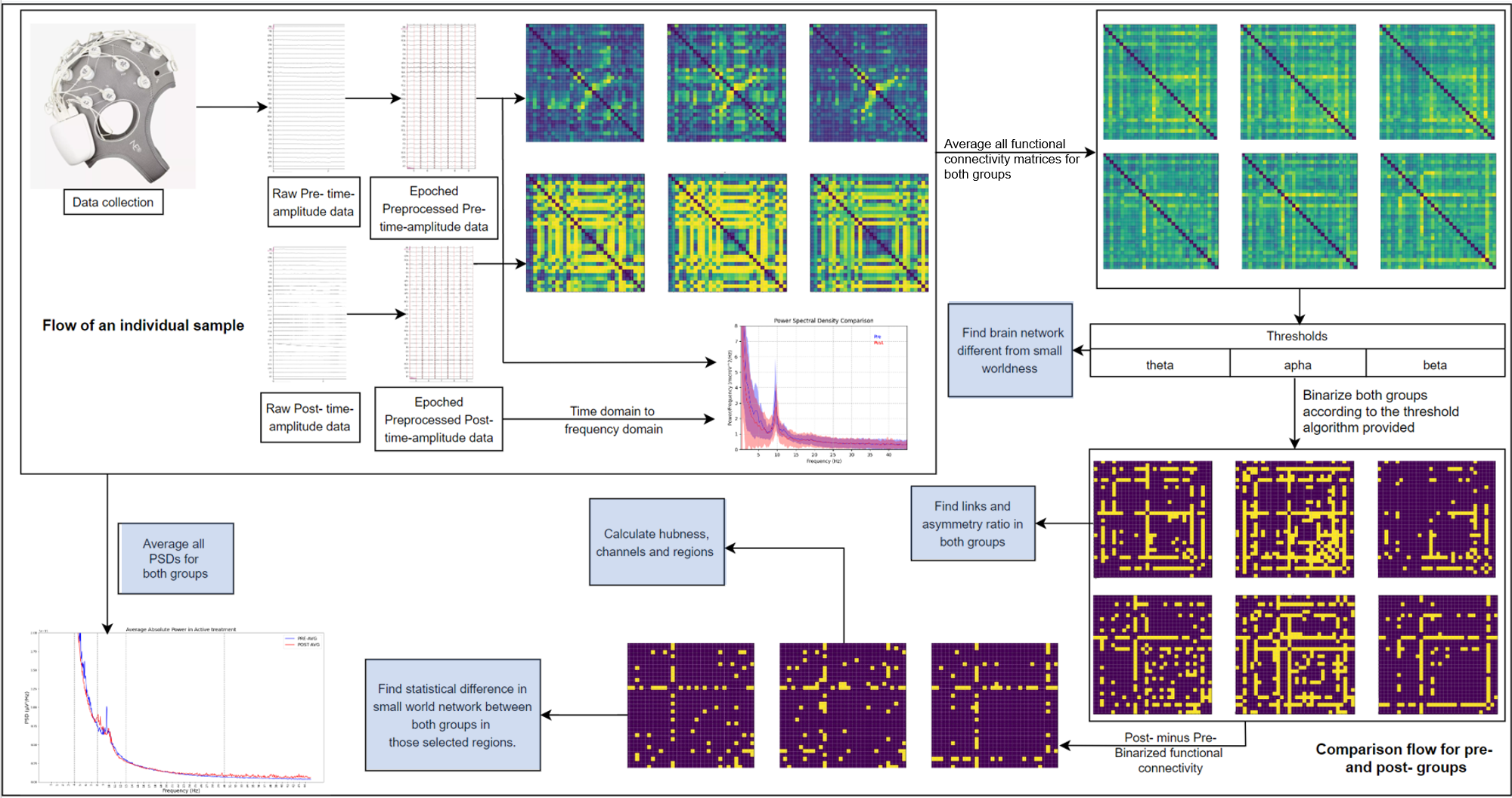}
    \caption{Flow of the analysis}
    \label{fig:flow}
\end{figure*}


\section{Tools \& Methods}
\begin{enumerate}{}{}
    \item \textbf{Pre-processing:} 
    Pre-processing of the EEG data involved several steps. First, all EEG data in both groups were average re-referenced to minimize the impact of common noise sources\cite{perrin1987practical}. Following this preprocessing, band-pass filtering was applied with a frequency range of $0.01-45$ Hz to isolate the EEG signals within the desired frequency band and to remove unwanted noise and artefacts. Then, the EEG data were cropped and epoched into segments with a duration of $4$ seconds each, with an overlapping window of $2$ seconds between consecutive epochs\cite{luck2014introduction}.

    \item \textbf{Power against frequency plot:} We plotted power against frequency for both pre-intervention and post-intervention groups, within the frequency range between $0.01-45 Hz$. 

    \item \textbf{Division of frequency bands and parcellation scheme}
    The division of frequency bands was as follows: theta ranges from $4-8 Hz$, alpha ranges from $8-13 Hz$ and beta ranges from $13-30 Hz$. We included Fp1, AF3, F3 and F7 in the left frontal region; Fp2, AF4, F4 and F8 in the right Frontal region; T7 in the left temporal region; T8 in the right temporal region; FC5, FC1, C3, CP1 and CP5 in the left central region; FC2, FC6, C4, CP2 and CP6 in the right central region; P7, P3, PO3 and O1 in the left parietal-occipital region; P4, P8, PO4 and O2 in the right parietal-occipital region.

    \item \textbf{Phase-based connectivity:} In phase-based connectivity, we used frequency domain features. Any signal may be analyzed in the frequency domain using the Fourier transform, which provides a frequency-domain representation of EEG data\cite{bracewell1986fourier}. However, the Fourier transform has some limitations\cite{cohen1995time}. Changes in frequency content in the signal over time are difficult to interpret using the Fourier transform, and it works well primarily for stationary signals. Since brain signals are dynamic and non-stationary, time-frequency decomposition is required to retain the advantages of both the time and frequency domains, though with some sacrifice in temporal and frequency precision\cite{mallat1999wavelet}. We used the Morlet wavelet (Gabor wavelet) for time-frequency analysis.
    
    \item \textbf{Phase lag index (PLI):} When two signals are perfectly in phase, they are synchronized. PLI works by focusing on the consistency of the phase differences between signals, rather than the absolute phase difference. With sensitivity towards the directionality of the phase difference between two signals, it measures the consistency of the phase lead/lag relationship between the signals, rather than just the magnitude of the difference. The fundamental idea is to disregard phase locking centred around 0 phase difference to exclude volume conduction effects (at the risk of ignoring true instantaneous interactions). This also applies to phase locking at $\pi, 2\pi$ and so on, i.e., repeating at every $\pi$, also given as 0 mod $\pi$. The PLI ranges between $0$ and $1$, with $0$ indicating no coupling due to volume conduction and $1$ indicating true, lagged interaction. We calculated it using multitaper which employs multiple windows to estimate the spectrum and thereby reduces the variance of the measurement. Mathematically it is expressed as follows\cite{stam2007phase}:

    \begin{equation}
    PLI = \left |   n^{-1}  \sum_{t=1}^{n}  sgn\left ( Im\left ( e^{i\left ( \phi^{j} - \phi^{k} \right )_{t}} \right ) \right )      \right |
    \tag{1}
    \end{equation}

    The vectors are not averaged with the PLI; instead, the sign of the imaginary part of the cross-spectral density is averaged. The PLI will be large if all phase angle variations are on one side of the imaginary axis. Conversely, the PLI will be zero if half of the phase angle discrepancies are positive and half are negative relative to the imaginary axis.
    
    \item \textbf{Small world network index (SWI):} The ``small-world'' phrase was derived from Stanley Milgram's 1967 study which demonstrated that people are connected through an average of six degrees of separation in social networks, revealing unexpectedly short paths between individuals\cite{milgram1967small}. Expanding upon this concept, Watts and Strogatz introduced the small-world network \cite{watts1998collective}. Before small-world networks, researchers typically modelled real-world networks using either random or regular networks. The concept of small-world networks revolutionized this understanding by showing that many real-world networks exhibit a small-world property, characterized by a high clustering coefficient like regular networks and a short average path length like random networks. Bassett and Bullmore implemented this model for brain anatomical and functional networks\cite{bassett2006small} demonstrating its capacity to support both segregated (specialized) and distributed (integrated) information processing. The SWI of a network is calculated as follows:
    
    \begin{equation*}
    \text{SWI} = \frac{C}{L_{\text{avg}}} \tag{2}
    \end{equation*}
    where \(C\) is the clustering coefficient of a network and \(L_{\text{avg}}\) is the average shortest path length of the network.
    
    \textit{Permutation comparison of random network SWI, pre-intervention SWI and post-intervention SWI}:
    We compared the SWI metric of the random network with that of the pre-intervention and post-intervention groups using the following algorithm:

    \renewcommand{\labelenumii}{\roman{enumii}.}
    \begin{enumerate}
        \item The algorithm iterates through a range of 200 thresholds (\(\Theta\)) from $0-1.5$ for each frequency band (\(\beta_f\)), where \(\beta_f \in \{\theta, \alpha, \beta\}\). The upper bound of $1.5$ is based on experimental observations indicating that values exceeding this range result in completely disconnected random networks.
        
        \item At each threshold (\(\Theta\)), we compute the threshold value (\(\Theta_{\text{thresh}}\)) for both pre-intervention and post-intervention groups (\(\Gamma\)), where \(\Gamma \in \{\text{pre}, \text{post}\}\) and for each frequency band \(\beta_f \in \{\theta, \alpha, \beta\}\). We define \(\Phi\) as the functional connectivity matrix, which depends on both the group and frequency band i.e., \(\Phi(\Gamma, \beta_f)\). Using \(\Phi\) and \(\Theta\), we calculate \(\Theta_{\text{thresh}}\):
    
        \begin{equation*}
        \Theta_{\text{thresh}} = \text{median}(\Phi) + \Theta \cdot \sigma(\Phi)
        \tag{3}
        \end{equation*}
        where, \(\Theta_{\text{thresh}}\) ranges for each connectivity matrix (\(\Phi\)) from \(\text{median}(\Phi)\) to \(\text{median}(\Phi) + 1.5 \cdot \sigma(\Phi)\).

        The different \(\Theta_{\text{thresh}}\) values for both treatment groups in \(\Gamma\) and in each \(\beta_f\) facilitate the calculation of the binary matrix (\(\mathcal{B}\)). This matrix is defined as a function of \(\Gamma\), \(\Theta_{\text{thresh}}\), and \(\beta_f\):
        \begin{equation*}
        \mathcal{B}(\Gamma, \beta_f, \Theta_{\text{thresh}}) = \Phi(\Gamma, \beta_f) > \Theta_{\text{thresh}} \tag{4}
        \end{equation*}
        Furthermore, we compute the number of connections (\(N_{\text{conn}}\)) and the SWI in \(\mathcal{B}(\Gamma, \beta_f, \Theta_{\text{thresh}})\).
    
        \item Firstly, for each \(\beta_f\) and \(\Theta_{\text{thresh}}\), we generate $100$ Erdős–Rényi random models (\(\mathcal{R}\)). The number of connections in these randomly created networks is set to the average \(N_{\text{conn}}\) of \(\mathcal{B}\) for both the groups in \(\Gamma \text{ i.e., } \{\text{pre}, \text{post}\}\). We then compute the SWI (\( \text{SWI}_{\text{random}}(\mathcal{R}_{\text{i}}) \)), clustering coefficient (\( \text{C}_{\text{random}}(\mathcal{R}_{\text{i}}) \)) and average path length (\( \text{L}_{\text{avg, random}}(\mathcal{R}_{\text{i}}) \)) for all $100$ random networks where \(i \in \{0, 1, 2, \ldots, 99\} \).

        Secondly, we randomly select two distinct networks \( \mathcal{R}_j \) and \( \mathcal{R}_k \) from the $100$ generated networks. The selected networks must exhibit some clustering (\(C \neq 0\)) and not have all nodes connected (\(L_{\text{avg}} \neq 0\)). We then calculate \(\text{SWI}_{\text{permutations}}(x)\) as follows:
        \begin{equation*}
        \text{SWI}_{\text{permutations}}(x) = \frac{\text{SWI}_{\text{random}}(\mathcal{R}j)} {\text{SWI}_{\text{random}}(\mathcal{R}_k)}
        \tag{5}
        \end{equation*}
        
        This process is repeated $1000$ times (\(x \in [0, 999]\)), generating a small-world network distribution of random networks centred around $1.0$. These $1000$ permutations represent random networks with equivalent topological properties but shuffled connections, creating a null distribution of small-world network values, \( \text{SWI}_{\text{permutations}}(x) \). We compare the observed small-world network value of the original networks for both groups against this null distribution. The observed real small-world network value is computed as:
            \begin{equation*}
            \text{SWI}_{\text{real}}(\Gamma, \beta_f) = \frac{\text{SWI}(\Gamma, \beta_f)} {(\frac{\text{\(\mu\)}(C_{\text{random}}(\mathcal{R}))}{\text{\(\mu\)}(L_{\text{avg, random}}(\mathcal{R}))})}
            \tag{6}
            \end{equation*}
        where \(\text{SWI}(\Gamma, \beta_f)\) is calculated on  \(\Phi(\Gamma, \beta_f)\). This comparison enables us to assess the significance of the network's small-world properties relative to random networks. Our threshold algorithm adapts a statistical approach for selecting a threshold based on normalized small-worldness\cite{10.7551/mitpress/9609.001.0001}. We conducted multiple experiments with this modified algorithm to optimize threshold selection by accounting for small-worldness in both pre-intervention and post-intervention groups.
    
        Finally, we compute the $p-value$ for both groups by comparing \( \text{SWI}_{\text{real}} \) against the \( \text{SWI}_{\text{permutations}} \) distribution.

        \item \textbf{Optimal threshold selection}: Z-scores are calculated for each \(\text{SWI}_{\text{real}}\) at each threshold for both groups to quantify the deviation of observed small-world network metrics from the null distribution. Ideally, we should select a threshold for which the small-world network metrics (\(\text{SWI}\)) are high for both groups and the difference between them is maximized. However, we have chosen a threshold for which the normalized small-world network metric exhibits the maximum difference between the groups (Figure \ref{fig:thresholdactive}).
    
    \end{enumerate}

   \item \textbf{Hubness:} We also implemented the concept of 'hub' or 'hubness' from graph theory in this section\cite{sporns2007brain}. A hub is a significant node in a network with numerous incoming and outgoing branches or connections to other nodes. In our case, a hub within an undirected network is characterized as a node with extensive connections to other nodes, with many nodes being indirectly linked through the hub. Consequently, if a hub node is removed, the network's functionality is significantly disturbed. Hubness implies a connectivity measurement indicating a node's potential to become a hub in a network. It is calculated for the difference matrix between the pre-intervention binarized group \(\mathcal{B}(\Gamma_{\text{pre}}, \beta_f, \Theta_{\text{thresh}})\) and the post-intervention binarized group \(\mathcal{B}(\Gamma_{\text{post}}, \beta_f, \Theta_{\text{thresh}})\). This involves counting the number of edges for each node, establishing a threshold to segregate channels with higher hubness values and identifying these channels as cortical regions of interest.

    \item \textbf{Asymmetry and the number of links:} We created a difference matrix by subtracting the pre-binarized matrix from the post-binarized matrix. This approach allows us to focus on the activity that increases post-intervention compared to pre-intervention in specific regions\cite{9288845}. The asymmetry ratio between the left and the right brain hemispheres is calculated by comparing the number of edges connecting to any left hemispheric nodes with those connecting to any right hemispheric nodes in the binary matrix obtained from the optimal threshold for each frequency band. We analyze the number of links after binarization to characterize the connectivity differences between both groups.

\end{enumerate}


\section{Results}

\subsection{\textbf{Sample characteristics: }}
Among the $12$ participants included in this study, approximately $67\%$ were female. The mean age of the sample population was $33.50 \pm 14.41$ years. The median illness duration was $18$ ($54$) months and the mean age of illness onset was $28.58 \pm 10.37$ years.

\subsection{\textbf{Changes in HAMD: }}
Mauchly's test for sphericity was non-significant. A repeated measures ANOVA (rmANOVA) determined that mean HAMD scores differed statistically significantly across the three time points ($F(2, 22) = 38.71$, $p<0.001$, $\eta^2=0.78$). Post hoc analysis with a Bonferroni adjustment revealed that the HAMD score was statistically significantly decreased from pre-intervention to post-intervention ($10.16$ ($95\%$ CI, $4.55-15.78$), $p=0.001$), and from pre-intervention to follow-up ($14.16$ ($95\%$ CI, $9.72-18.61$), $p<0.001$), and from post-intervention to follow-up ($4.01$ ($95\%$ CI, $0.19-7.80$), $p=0.03$), which indicates the beneficial effect of tDCS intervention for these individuals.

\subsection{\textbf{Power spectral density: }}
\begin{figure}[htbp]
    \centering
    \includegraphics[width=0.485\textwidth]{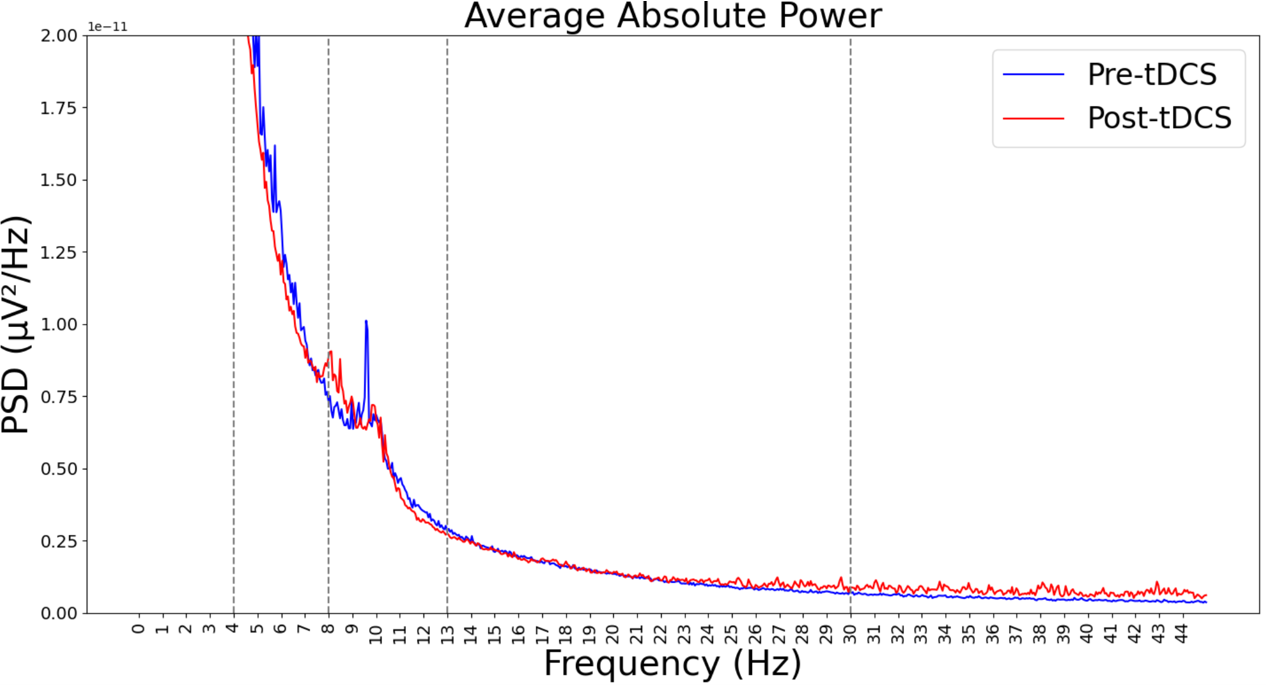}
    \caption{Power spectral density (PSD) plot for pre-intervention (blue) vs post-intervention (red)}
    \label{fig:psdactive}
\end{figure}

We calculated Power spectral density (PSD) using the Welch method for both pre-intervention and post-intervention for theta, alpha, and beta bands (Figure \ref{fig:psdactive}). It is observed that the alpha band gets reduced in the post-intervention compared to the pre-intervention group.

\subsection{\textbf{Functional connectivity in frequency bands: }}
\begin{figure}[htbp]
  \centering
  \includegraphics[width=\linewidth]{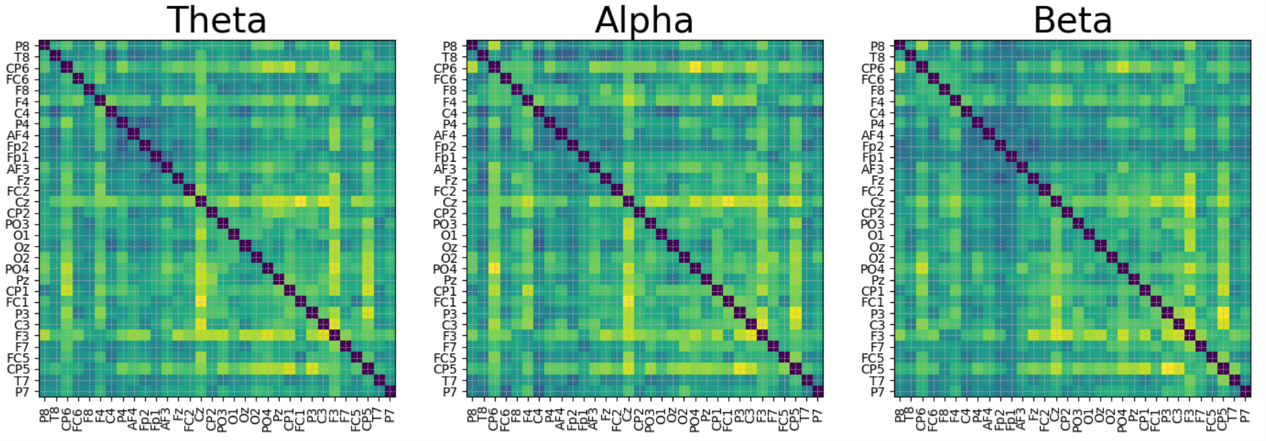}
  
  \includegraphics[width=\linewidth]{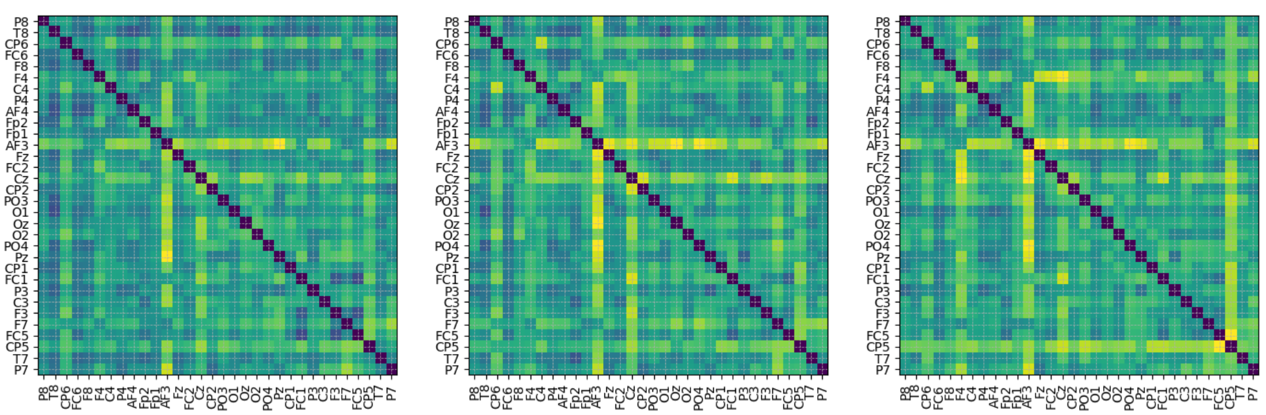}
    \caption{Functional connectivity for a) Pre-intervention b) Post-intervention in all theta, alpha and beta bands for tDCS treatment}
    \label{fig:fcprepostactive}
\end{figure}

We calculated functional connectivity using PLI with the multitaper method in both pre-intervention and post-intervention groups. We observed that certain electrodes exhibited more activity in the pre-intervention group  (Figure \ref{fig:fcprepostactive}). These electrodes included CP6, F4, Cz, F3, and CP5 across all frequency bands. In the post-intervention group, a different set of electrodes showed increased connectivity. These included CP6, AF3, Cz, and CP5 in the theta band; CP6, AF3, Cz, CP5, and F7 in the alpha band; and F4, AF3, and CP5 in the beta band.

\subsection{\textbf{Optimal threshold to binarize functional connectivity: }} 
\begin{figure}[htbp]
  \centering
  \includegraphics[width=\linewidth]{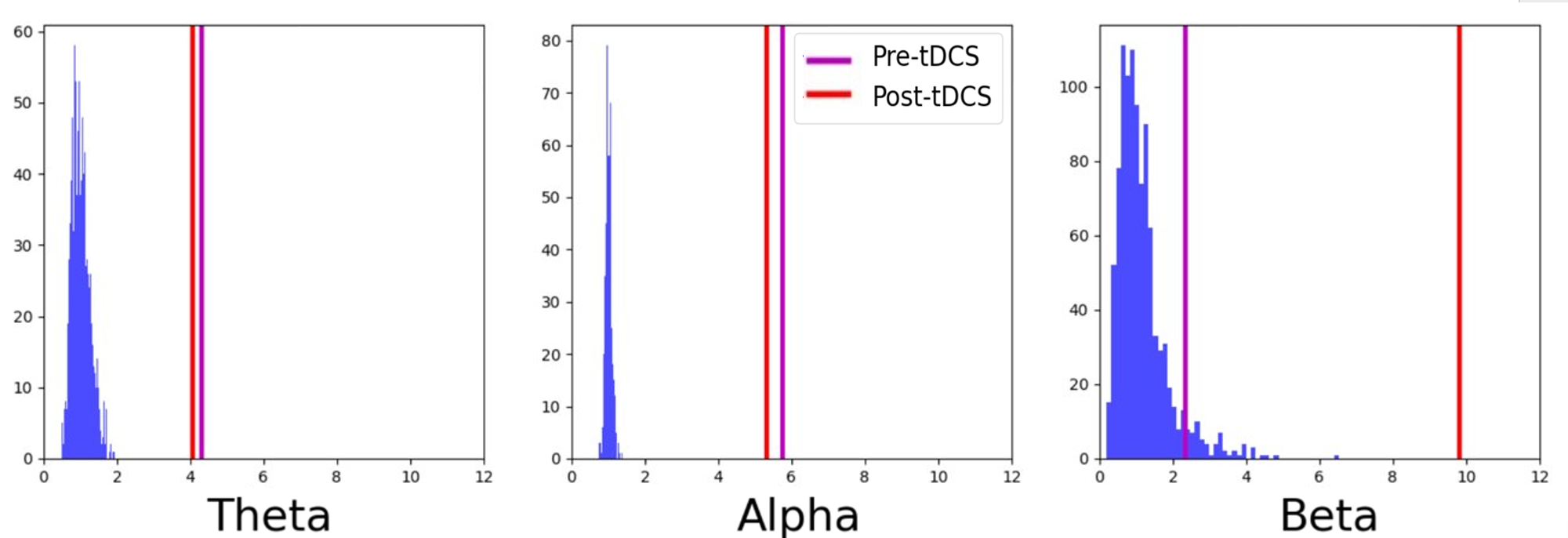}
    \caption{Comparison of random network SWI distribution, pre-intervention SWI and post-intervention SWI plots at optimal threshold for each frequency band}
    \label{fig:randomactive}
\end{figure}

We calculated the threshold using the algorithm provided above. Subsequently, we calculated the random network distribution for all frequency bands and plotted it against the small-world network for both pre-intervention and post-intervention groups. For the theta and alpha bands, we observed that the p-value for both pre-intervention and post-intervention groups was almost equal to $0.0$. However, in the beta band, we observed that the p-value for the pre-intervention group compared to the random network distribution was $0.063$, whereas, for the post-intervention group, it was $0.001$ (Figure \ref{fig:randomactive}).


\begin{figure*}[!t]
  \centering
  \subfloat[Theta]{\includegraphics[width=0.33\linewidth]{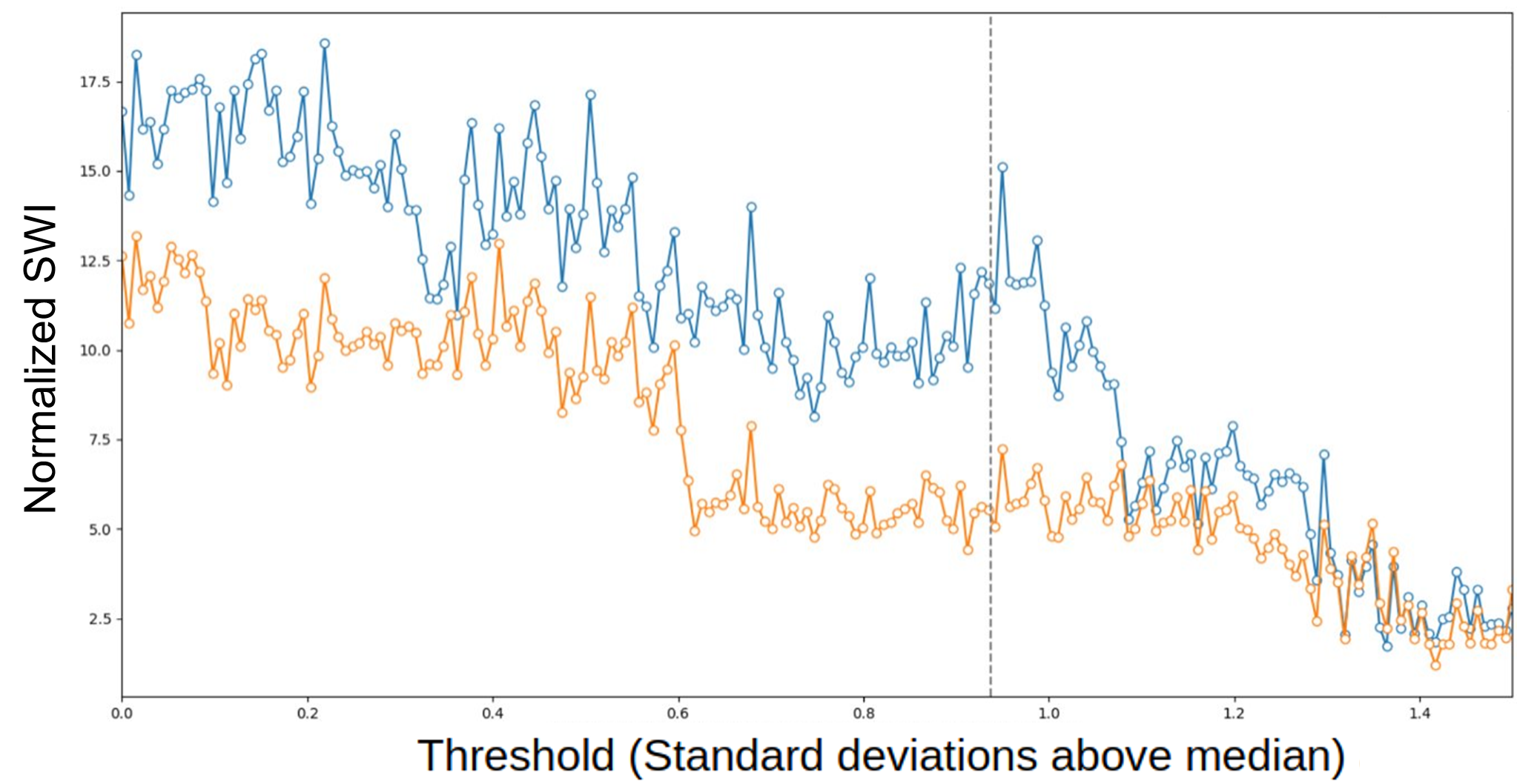}}
  \subfloat[Alpha]{\includegraphics[width=0.33\linewidth]{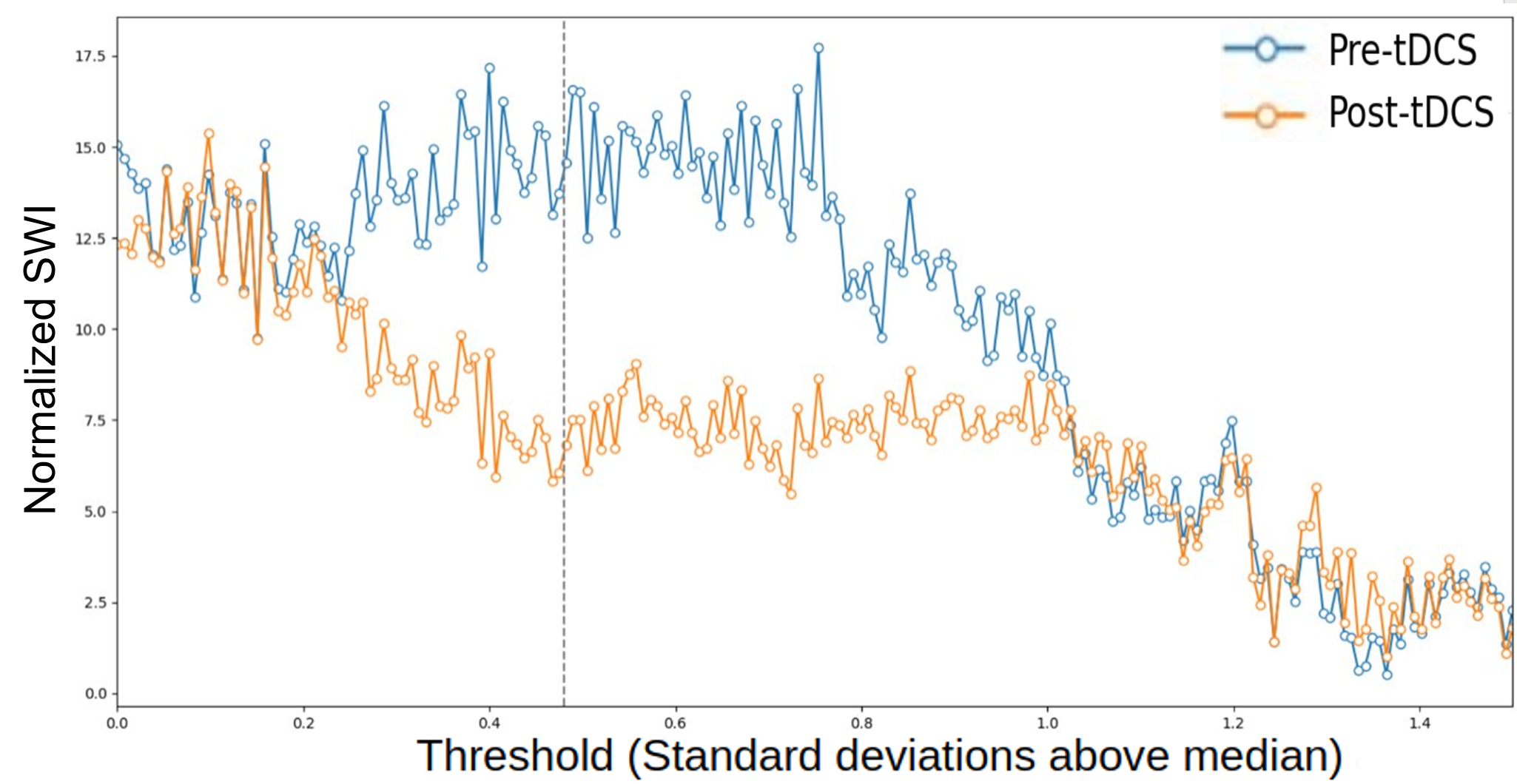}}
  \subfloat[Beta]{\includegraphics[width=0.33\linewidth]{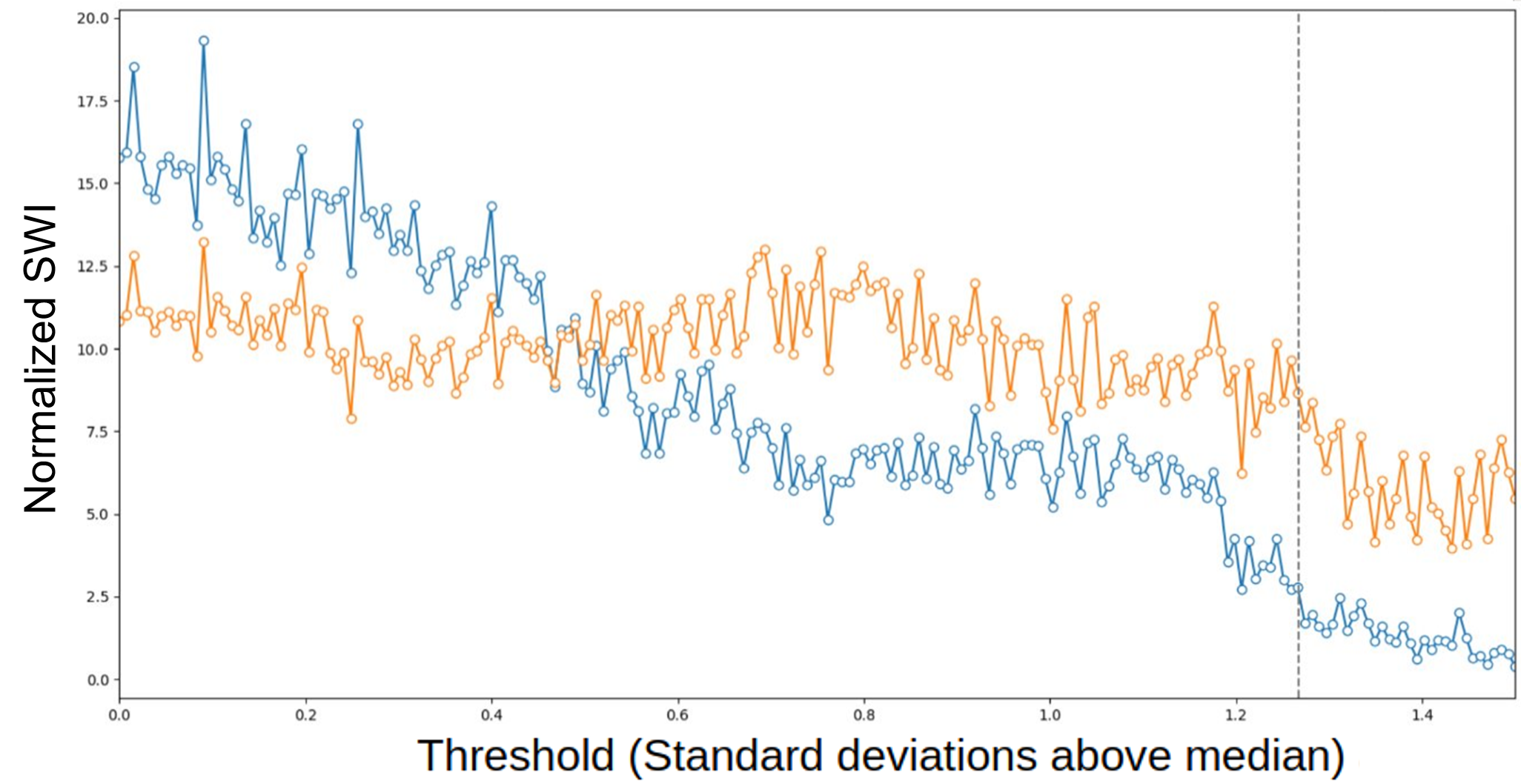}}
    \caption{Optimal Threshold selection for all frequency bands in tDCS Treatment}
    \label{fig:thresholdactive}
\end{figure*}

We maximized the difference between the z-normalized small-world networks of the two groups and determined the optimal thresholds for each frequency band (Figure \ref{fig:thresholdactive}). 

\subsection{\textbf{Regions of brain selected based on hubness of post minus pre-intervention binarized matrices}}

\begin{figure*}[!t]
  \centering
  \includegraphics[width=1\linewidth]{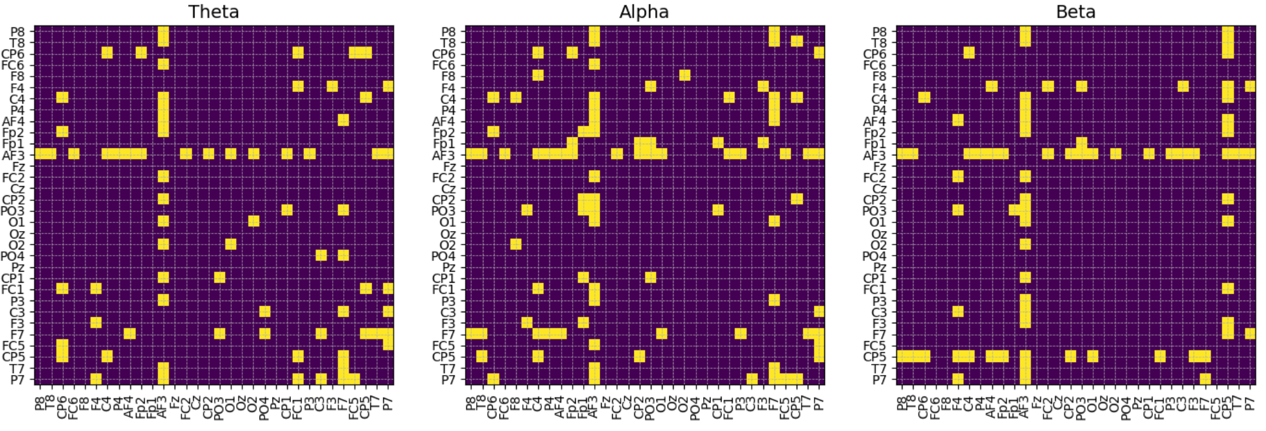}
    \caption{Binarized functional activity in post-intervention minus pre-intervention across all frequency bands}
    \label{fig:binarizedactive}
\end{figure*}

We observed that different hubness thresholds yielded different brain regions of interest. Following the parcellation method described above, we identified 3-4 channels in each frequency band, encompassing LF, RC and LPO regions for theta and alpha bands at a hubness threshold  \( \geq \) $5$ edges and \( \geq \) $6$ edges respectively. For the beta band, we found regions in the RF, LF and LC at a hubness threshold \( \geq \) $6$ edges (Figure \ref{fig:binarizedactive}). 

\subsection{\textbf{Links and asymmetry ratio in both pre- and post-intervention binary matrix }}
\begin{figure}[htbp]
  \centering
  {\includegraphics[width=1\linewidth]{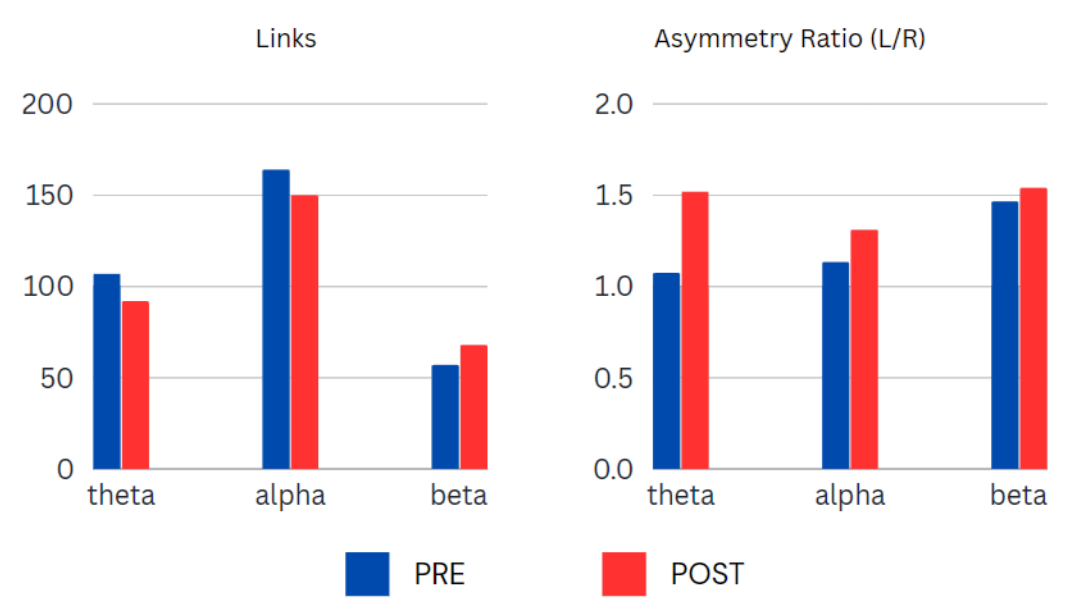}}
    \caption{Links and asymmetry ratio for both pre-intervention and post-intervention in all frequency bands}
    \label{fig:linksasym}
\end{figure}

We compared the links and asymmetry ratio for both pre-intervention and post-intervention groups. The results showed a decreased number of links in theta and alpha bands, but an increase in the beta band. Additionally, the asymmetry ratio increased across all frequency bands. An increased asymmetry ratio indicates more activity in the left hemisphere compared to the right (Figure \ref{fig:linksasym}).

To integrate the discussion provided into the existing context and compare our results with related works, we can expand upon the findings and implications of our study.

\section{Discussion} 

\subsection{\textbf{Depressive symptoms in MDD are reduced after tDCS}}
Similar to our findings, recent literature has reported a significant reduction in MDD symptoms following tDCS intervention \cite{mcgirr2018clinical, mutz2019comparative, fregni2021evidence}. Previous work from our lab also demonstrated significant improvement in symptoms after tDCS intervention\cite{verma2024effectiveness}.

\subsection{\textbf{Overall brain alpha activity is reduced after tDCS}}
Jernajczyk ($2017$) observed that the difference in alpha wavelet power between responders and non-responders to antidepressant treatment was most pronounced in the occipital channels \cite{jernajczyk2017alpha}. Segrave ($2011$) also identified lateralized differences in frontal alpha power in individuals with MDD, suggesting an aberrant affective processing style \cite{segrave2011individualized}. This observation was further supported by Ischebeck ($2014$), who found altered frontal EEG asymmetry in obsessive-compulsive disorder, a condition often associated with depression \cite{ischebeck2014altered}. Our finding of reduced alpha band activity following tDCS treatment is consistent with previous research indicating that modulation of neural oscillatory patterns is associated with mood regulation \cite{smith2015transcranial}. This reduction in alpha activity suggests a neurophysiological response to tDCS intervention that may reflect its therapeutic effects.

\subsection{\textbf{Average PLI value in DLFPC region is more after tDCS}}
Our study reveals that, before the tDCS intervention, certain brain electrode locations, particularly in the central regions, exhibited higher connectivity. However, post-intervention, the dorsolateral prefrontal cortex (DLPFC) regions demonstrated increased connectivity. This finding aligns with previous literature, which will be further discussed in the subsequent section, highlighting how connectivity links were more numerous but dispersed across the overall brain in the theta and alpha bands compared to after tDCS. The identification of specific hub regions such as the left central and the left frontal areas, exhibiting increased network centrality following tDCS intervention, supports findings from studies that implicate frontal brain regions in tDCS-mediated antidepressant effects\cite{voros2018direct}. For instance, Olbrich et al. ($2014$) and Jiang et al. ($2019$) reported increased connectivity in the frontal and central regions in individuals with MDD. Olbrich specifically noted enhanced prefrontal connectivity at the alpha frequency, while Jiang identified elevated frontolimbic and frontocentral connectivity mediated by gamma activity\cite{olbrich2014functional, jiang2019hyperactive}. These results highlight the neural activity patterns in specific brain circuits associated with mood regulation linked to the response to tDCS.

\subsection{\textbf{Overactivation of connectivity in the theta and the alpha band before tDCS}}
For the overall brain region, we observed a change in the connectivity number with a decrease in connectivity number for theta and alpha bands and an increase in beta bands after tDCS intervention as compared to baseline suggesting that the alert state became more prominent and relaxed state became less prominent after tDCS in our study. The literature supports this observation, indicating that alpha and theta bands are effective discriminators between depressed individuals and healthy controls, relating these bands to emotional processing\cite{aftanas2001human, aftanas2001event, hosseinifard2013classifying}. Hasanzadeh et al. ($2017$) reported higher average PLI in the MDD group across alpha, beta and total frequency bands\cite{hasanzadeh2017investigation}. Moreover, Fogelson et al. ($2020$) also reported increased cluster coefficient and local efficiency during predictive stimulus processing in MDD, suggesting overactivation in frontal networks\cite{fogelson2020functional}.

\subsection{\textbf{Randomisation of brain network before tDCS in the beta band}}
Our analysis of binarized functional connectivity networks revealed significant alterations in network topology, particularly in the beta band, indicative of enhanced connectivity and functional reorganization associated with tDCS treatment. This observation aligns with previous studies on the effects of tDCS on brain networks, highlighting potential mechanisms underlying therapeutic responses \cite{liston2014default}. Li et al. ($2015$) reported that patients with depression exhibited a more randomized brain network during emotional face processing\cite{li2015abnormal}. Similarly, Liu et al. ($2022$) documented a more randomized network structure in MDD individuals across the delta, theta, alpha and beta bands\cite{liu2022hypofunction}. Hazandeh et al. ($2020$) observed that MDD patients displayed a more randomized network structure with disrupted directed interactions between various brain regions and between information in the left and right hemispheres\cite{hasanzadeh2020graph}.

\subsection{\textbf{Observed left asymmetry across all frequency bands post tDCS}}
Research highlights that brain asymmetry is a fundamental characteristic of a healthy brain, with a frontal-occipital gradient of cortical thickness asymmetry being a notable finding\cite{kong2022mapping}. This asymmetry is also evident in the structural connectivity of healthy older adults, with leftward asymmetry in medial temporal, dorsolateral frontal, and occipital regions, and rightward asymmetry in middle temporal and orbitofrontal regions\cite{bonilha2014asymmetry}. The hemispheric network, which delineates structural and functional connectivity within each hemisphere is a crucial framework for investigating brain asymmetry and its implications in health and disease\cite{wang2023brain}. The genetic architecture of this structural left-right asymmetry has also been explored, with significant loci associated with brain asymmetry identified\cite{sha2021genetic}. These findings suggest that though left asymmetry is more common in healthy brains, it might be altered in certain conditions. Our study observed the increased left asymmetry ratio in all the frequency bands following tDCS.

\subsection{\textbf{Limitations and drawbacks}}
\begin{enumerate}{}{}
    \item \textbf{Sample size: }The study was conducted with a small sample size of twelve participants, which limits the generalizability of the findings to broader populations.
    \item \textbf{Data variability: }The small sample size introduces variability in the data, including differences in illness duration and gender. Additionally, the severity of depression may introduce biases in the study outcomes.
    \item \textbf{Threshold dependency: }The analysis involved binarizing functional connectivity matrices from pre-intervention and post-intervention data, which inherently depends on the chosen threshold. Because of this dependency, we have invested significant effort and time to develop a robust algorithm. It is statistically derived and experimented with under different circumstances. The number of connections (edges) in the random network distributions was selected to match the average number of connections observed in both groups. The threshold value is inherently variable but will typically cluster around the value we obtained, as the process generates random networks at each iteration while maintaining a constant number of edges. So, re-running the whole flow might produce slightly different results in the number of links, asymmetry, channels, and hubness values but will differ only minimally from the results presented in this paper.
\end{enumerate}

In summary, our study provides comprehensive evidence supporting the efficiency of tDCS as a non-invasive neuromodulation technique for modulating brain network properties in individuals with MDD. It extends previous research by uniquely focusing on modifying binarizing thresholding algorithms to analyze functional connectivity networks in the context of tDCS for MDD. The results align with the prior findings demonstrating tDCS-induced alterations in neural activity and network properties across various frequency bands\cite{kadosh2016changes, liston2014default}. By refining analysis techniques and integrating insights from previous studies, our work provides new perspectives on tDCS’s impact on brain connectivity dynamics, offering valuable contributions to the therapeutic strategies for MDD.

\section{Conclusion and future works}

In this study, we evaluated the efficiency of tDCS for MDD by analyzing resting-state EEG data and assessing functional connectivity using network neuroscience across different frequency bands. A novel aspect of our research was modifying the binarizing thresholding algorithms to provide a detailed understanding of network topology changes following tDCS treatment. Despite these advancements, limitations such as a small sample size, data variability, and reliance on thresholding methods must be addressed. Future research with larger sample sizes and longitudinal designs is necessary to confirm these findings and explore the long-term effects of tDCS on brain connectivity in MDD.

Our findings contribute to the literature on the neural mechanisms of MDD and the potential for tDCS as a therapeutic intervention. We observed significant changes in brain network randomization in the beta band post-tDCS. Additionally, differences in PSD in the alpha band were noted between pre-intervention and post-intervention groups. Moreover, an increased number of connectivity links in the theta and alpha bands were found pre-intervention which were more scattered but became more localised in the DLFPC region in all the bands. Furthermore, left hemispheric asymmetry increased following tDCS. These insights could inform the development of personalized treatment strategies for MDD and other psychiatric conditions. Overall, while tDCS shows promise as a non-invasive treatment for MDD, further research is needed to validate and refine its clinical application.

\textbf{Funding:} This research received no external funding.
\\
\textbf{Conflict of interest:} The authors declare no conflict of interest.

\section{Acknowledgement}
The authors thank all the twelve participants who facilitated the analysis of the tDCS intervention and the technicians for the extensive data collection procedure. We acknowledge using Elicit: Notebook\footnote{\url{https://elicit.com/notebook/}} to assist in generating relevant research papers.

\bibliographystyle{IEEEtran}
\bibliography{MDD}

\vfill
\end{document}